\documentclass{aa}
\usepackage{amsfonts}
\usepackage{amsmath}
\usepackage{amssymb}
\usepackage{graphicx}
\usepackage{natbib}
\usepackage{txfonts}
\bibpunct{(}{)}{;}{a}{}{,}
\begin{document}
\title{Consistency between deep crustal heating of strange stars in
  superbursters and soft X-ray transients} 
\titlerunning{Deep crustal heating of strange stars in superbursters
  and soft X-ray transients}
\authorrunning{M. Stejner \and J. Madsen}

\author{Morten Stejner \and
  Jes Madsen} 

\institute{Department of Physics and Astronomy,
  University of Aarhus, DK-8000 \AA rhus C, Denmark\\
  \email{msp@phys.au.dk, jesm@phys.au.dk}}

\date{Received / Accepted}

\abstract{Both superbursters and soft X-ray transients probe the
  process of deep crustal heating in compact stars. It was recently
  shown that the transfer of matter from crust to core in a strange
  star can heat the crust and ignite superbursts provided certain
  constraints on the strange quark matter equation of state are
  fulfilled.}
{We derive corresponding constraints on the equation of state for
  soft X-ray transients assuming their quiescent emission is powered
  in the same way, and further discuss the time dependence of this
  heating mechanism in transient systems. }
{We approach this using a simple parametrized model for deep crustal
  heating in strange stars assuming slow neutrino cooling in the core and
  blackbody photon emission from the surface.}
{The constraints derived for hot frequently accreting soft X-ray
  transients are always consistent with those for superbursters. The
  colder sources are consistent for low values of the quark
  matter binding energy, heat conductivity and neutrino
  emissivity. The heating mechanism is very time dependent which may
  help explain cold sources with long recurrence times. Thus deep
  crustal heating in strange stars can provide a consistent
  explanation for superbursters and soft X-ray transients.}
{} 


\maketitle
\section{Introduction}
Strange stars are a class of compact stars made entirely of absolutely
stable strange quark matter except for a possible thin crust of
ordinary nuclear matter below the neutron drip density separated from
the core by a strong Coulomb barrier
\citep{Itoh:1970,Baym:1976,Witten:1984,Haensel:1986,Alcock:1986}. The stability
of strange quark matter depends on poorly constrained strong
interaction properties and remains to be decided by observation or
experiment (see \cite{Madsen:1999,Weber:2005,Weber:2006} for
reviews). If true, the strange matter hypothesis would imply that all
compact stars are likely to be strange stars (because the galactic flux of quark matter lumps from strange star
collisions will trigger the hadron-to-quark transition in neutron stars
and supernova cores~\citep{Madsen:1988,Caldwell:1991}), and so strange star
models should be able to reproduce observed characteristics of compact
stars. Specifically strange stars should be able to reproduce the
bursting behavior of low mass X-ray binaries (LMXBs) expected to
contain some of the heaviest compact stars due to accretion, and here
we test the consistency of the recent model for superburst ignition in
strange stars by \cite{Page:2005} with similar considerations for the
quiescent emission from soft X-ray transients; two of the most
important LMXB activities.

Superbursts, first discovered by \cite{Cornelisse:2000}, are a rare
kind of type I X-ray bursts from LMXBs distinguishing themselves by
extreme energies ($\sim 10^{42}$ erg), long durations (4--14 hours)
and long recurrence times ($\sim 1 $ year) -- see \cite{Kuulkers:2004}
for a recent review of their observational properties. The mechanism
behind superbursts is thought to be unstable thermonuclear burning of
carbon in the crust at densities around $10^{9}\mbox{ g cm}^{-3}$
\citep{Cumming:2001,Strohmayer:2002,Cumming:2004,Brown:2004,in'tZand:2004,Cooper:2005,Cumming:2005}. The
fuel for such an event is provided by the burning of accreted hydrogen
and helium at the surface, which may take place either stably or
unstably. In the latter case it gives rise to ordinary type I X-ray
bursts. The ashes from this process, containing about 10\% carbon by
mass, are compressed under the increasing weight of the layers above,
and when enough carbon has been accumulated it may be ignited and burn
unstably. Observed superburst light curves require the ignition to
take place at temperatures around $6\times 10^8$ K and column depths
of approximately $10^{12} \mbox{ g cm}^{-2}$, and as the heat released by
accretion and burning at the surface is quickly reradiated and does
not penetrate to this depth, these ignition conditions must be set by
the process of deep crustal heating and therefore probe the physics of
the deep crust and core \citep{Brown:2004,Cooper:2005}.

In neutron stars deep crustal heating refers to the heat released as
the sinking ashes undergo electron captures and most importantly -- at
densities above $10^{12} \mbox{ g cm}^{-3} $ -- pycnonuclear reactions
\citep{Haensel:1990,Haensel:2003,Brown:1998}. This amounts to about 1.4 MeV per nucleon and heats the
core to $\sim 10^8$ K depending on the accretion rate. It has however
been difficult to achieve superburst ignition at the appropriate
column depth, and in a recent investigation \cite{Cumming:2005} showed
that neutrino emission from Cooper pairing of neutrons at densities in
the range $10^{12}-10^{13} \mbox{ g cm}^{-3}$ limits the crust
temperature below $5\times 10^8$ K. Thus unstable carbon ignition
seems impossible at the observed depth unless Cooper pair emission is
less efficient than currently thought or some additional heating
source can be found.

In strange stars there can be no pycnonuclear reactions because the
crust density is limited by the neutron drip density, but for the same
reason there can be no Cooper pair emission from neutrons
either. Furthermore the energy released by matter tunneling through the
Coulomb barrier at the base of the crust and converting to quark
matter could be much greater than that released by pycnonuclear
reactions in neutron stars. Thus it seems reasonable to expect that
strange stars in LMXBs can be heated to higher temperatures than
neutron stars, and \cite{Page:2005} recently showed that this can lead
to superburst ignition at the observed column depths if certain
constraints on the strange quark matter core parameters are fulfilled.

Soft X-ray transients are LMXBs which undergo accretion outbursts
lasting days to years separated by long periods of quiescence in which
no or very little accretion takes place and the X-ray luminosity drops to
$\sim 10^{32} \mbox{ erg s}^{-1}$
\citep{Campana:1998,Rutledge:1999,Tomsick:2004}. When accretion stops,
the heat deposited at the surface quickly radiates away, and so in the
absence of any residual accretion the surface temperature should
reflect the temperature of the core including the effects of deep
crustal heating. This has been thoroughly investigated for neutron
stars in the literature \citep[ and references
therein]{Brown:1998,Ushomirsky:2001,Colpi:2001,Brown:2002,Rutledge:2002,Yakovlev:2003,Yakovlev:2004}
establishing deep crustal heating as an unavoidable heating mechanism
in the core, which should at least provide a ``rock bottom''
luminosity for these sources. Residual accretion may still play a role
however -- indeed the power law components in the spectra of many of
these sources are often taken as a sign of residual accretion or shock
emission from a rotation powered pulsar -- and furthermore the
properties of the core remains uncertain, so the precise nature of the
quiescent emission is not firmly established.

Considering the many possible phases and poorly constrained properties
of strange quark matter one would expect -- for the same general
reasons they were able to fit superbursters -- that a wide range of
strange star models can reproduce the observed properties of soft
X-ray transients. Our approach is therefore to use a simple
parametrized model for deep crustal heating in strange stars to
discuss the consistency between constraints on strange quark matter
derived from superbursters and soft X-ray transients, as such a mutual
fit is likely to constrain the strange quark matter equation of state further.

We describe the overall structure of strange stars and their crusts in
Sect.~\ref{sec2} and then turn to deep crustal heating in Sect.
\ref{sec3}. In Sect.~\ref{sec4} we compare models with different
assumptions about the cooling mechanism to observations
of soft X-ray transients, and we discuss the importance of time
dependence in Sect.~\ref{sec5} before concluding in Sect.
\ref{sec6}.

\section{Strange stars and their crusts}\label{sec2}
If strange stars are stable they contain roughly equal numbers of up,
down and strange quarks in their core, but due to the higher mass of
the negative s-quark, these will normally be slightly disfavored
giving the core an overall positive quark charge to be compensated by
electrons. Even color-flavor locked strange quark matter, which is
strictly charge neutral in bulk \citep{Rajagopal:2001} has an overall
positive quark charge due to surface effects
\citep{Madsen:2000,Madsen:2001,Usov:2004}. Because strange quark
matter is self bound the density is largely constant in the core
dropping by only a factor of a few from the center to the surface,
where it goes from above nuclear density to zero within a few fm --
see
e.g. \cite{Alcock:1986,Haensel:1986,Glendenning:1992,Kettner:1995,Weber:1999,Glendenning:2000,Zdunik:2001,Zdunik:2002}
for details on the global properties of strange stars. Since the
electrostatic force is much weaker than the strong force confining the
quarks, some of the electrons necessary to create an overall charge
neutral object will form a thin atmosphere outside the quark phase
held in place by a huge electric field. The electric field -- up to
$10^{18} \mbox{ V cm}^{-1}$ -- is capable of sustaining a nuclear matter
crust decoupled from the quark phase by a small electron filled gap
\citep{Alcock:1986,Kettner:1995,Stejner:2005}. The crust is
electrically neutral in bulk however, so beyond the boundary layer its
structure is given by hydrostatic equilibrium with an equation of
state corresponding to the chemical composition of the ashes from the
surface burning \citep{Glendenning:1992,Zdunik:2001}. The crust mass
roughly follows \citep{Zdunik:2001}
\begin{align}
M_\mathrm{C}&=\frac{4\pi R^4 P_\mathrm{b}}{GM}\left(1-\frac{2GM}{Rc^2}\right)\\
&=8.1\times
10^{-5}\frac{R_6^4}{M_*}\left(\frac{\rho_\mathrm{b}}{\rho_\mathrm{D}}\right)^{\gamma}\left[1-0.295\frac{M_*}{R_6}\right]
M_\odot \label{mass}
\end{align} 
where $P_b$ is the pressure at the bottom of the crust,
$M_*=M/M_\odot$ is the stellar mass in units of the solar mass,
$R_6=R/ 10\mbox{ km}$ is the stellar radius, $\rho_\mathrm{b}$ is the
density at the bottom of the crust, $\rho_\mathrm{D}=7.8 \times
10^{11} \mbox{ g cm}^{-3}$ is the neutron drip density, $\gamma\simeq
4/3$ is the adiabatic index in the crust and the last equality assumes
the crust composition found by \cite{Haensel:2003} for accreting
neutron stars. The density at the bottom of the crust may be limited
by the neutron drip density if the Coulomb barrier from the core is
sufficiently high, which depends on the details of the equation of
state of strange quark matter. In most cases this will not be possible
though and the maximum crust density is then determined by the rate at
which ions tunnel through the Coulomb barrier such that the tunneling
rate for an accreting star is equal to the accretion rate or
corresponds to the inverse stellar age if
the star is isolated (see \cite{Stejner:2005} for a detailed
discussion). The tunneling rate close to this equilibrium is very
sensitive to the crust density so there is no great difference between
these two cases in terms of crust density, but for the same reason one
may expect that in soft X-ray transients the tunneling rate will vary
with the accretion rate -- we return to this point in
Sect.~\ref{sec5}.

Since the crust has no significant heating sources of its own
(electron captures in  the crust amount to less than 40 keV per
nucleon \citep{Haensel:2003}) its thermal profile is set by the
diffusion of heat from the core towards the surface. The surface
temperature, $T_\mathrm{S}$, is therefore only a function of the core
surface temperature, $T_\mathrm{i}$, and thus probes the process of
deep crustal heating. A detailed account for such
$T_\mathrm{S}-T_\mathrm{i}$ relations can be found in
e.g. \cite{Yakovlev:2004}, but in a reasonable approximation
$T_\mathrm{i,9} \simeq 0.1T^2_{S,6}$ \citep{Gudmundsson:1983}, which
will be sufficient for our purpose. Here
$T_\mathrm{S,6}=T_\mathrm{S}/10^6$ K and
$T_\mathrm{i,9}=T_\mathrm{i}/10^9$ K is actually the temperature at a
density of $10^{10} \mbox{ g cm}^{-3}$, but as the crust is largely
isothermal at this depth, one can take this to be the core surface
temperature \citep{Gudmundsson:1983,Yakovlev:2004}.

In the discussion above and in what follows we have assumed the
conventional description of strange stars with a sharp core surface,
an electron filled gap and a nuclear crust. It should be
mentioned however that \cite{Jaikumar:2006} recently showed, that if the
surface tension of strange quark matter is not too high, it may be
unstable to phase separation at low pressure leading to a crust of
quark nuggets embedded in a degenerate electron gas, which would in
many ways resemble the nuclear crust on neutron stars. The consequences
of such a crust in an accreting system are not clear at present, and
we restrict ourselves to the traditional scenario.  

\section{Deep crustal heating in accreting strange stars} \label{sec3}
Deep crustal heating in strange stars was recently studied by
\cite{Page:2005} as a means of achieving superburst ignition conditions
at the appropriate column depth and thus constrain the properties of
strange quark matter (see also the work on continuously accreting
strange stars by \cite{Escude:1990}). It was shown that observed
ignition conditions are achieved for a wide range of crust density,
$\rho_\mathrm{b}$, strange quark matter binding energy, $Q_\mathrm{SQM}$, neutrino
emissivity, $\epsilon_\nu$, and thermal conductivity, $K$, provided
that the direct Urca process does not operate in the core implying
that strange quark matter should be a color
superconductor. Furthermore a set of useful analytical approximations
to the core surface temperature were shown to reasonably approximate
numerical calculations. Here we will use those same approximations to
find the corresponding constraints on strange quark matter in soft
X-ray transients and a relationship between the observable parameters
of superbursters and soft X-ray transients. For clarity and to
emphasize the approximations we therefore briefly review the
calculation by \cite{Page:2005} in the following subsection.
\subsection{Approximate thermal structure}
As accretion on the strange star drives the crust towards its maximum
density, the rate at which nuclei (or neutrons if the neutron drip
density can be reached) tunnel through the Coulomb barrier into the
quark matter core will increase until an equilibrium crust density is
reached. This density may vary as a function of the unknown Coulomb
barrier height at the quark matter surface but this would not change
the qualitative features of the model and for our purpose we will
assume that the equilibrium crust density is $\sim 10^{11}\mbox{
g cm}^{-3}$. As the nuclei cross into the strange quark matter core at
the rate $\dot{m}$ (in units of nucleons cm$^{-2}$ s$^{-1}$
unless otherwise stated) and are dissolved, they release an energy,
$Q_\mathrm{SQM}\leq 100$ MeV per nucleon (the actual value depends on
uncertain values for strong interaction parameters but is fixed if 
these parameters were known). This heats the core until
an equilibrium or steady state temperature is reached as the heating
balances energy loss by neutrino emission in the core. This is assumed
to take place at a (slow) rate of
\begin{align}\label{epsilonnu}
\epsilon_\nu\simeq Q_\nu\times T^8_9 \mbox{ erg cm}^{-3}\mbox{ s}^{-1},
\quad Q_\nu \sim 10^{18}-10^{22}
\end{align} 
with $T_{9}=T/10^9 \mbox{ K}$, corresponding to modified Urca and
bremsstrahlung processes \citep{Page:2005b,Yakovlev:2004b}. We thus
neglect a slow dependence on density in $Q_\nu$ which would only
amount to a factor of order unity since the core density is nearly
constant.

We will -- at first -- assume that this equilibrium is simply
determined by the average accretion rate, which would be reasonable in
systems where the tunneling is more or less continuous. We return
to this point in section~\ref{sec5} however as this steady state
assumption may not be sufficient for all systems.

Following \cite{Page:2005} we model the thermal profile
of the core in a plane parallel approximation with heat transport
taking place by conduction and with a heat sink from neutrino
emission. The inward directed heat flux from the conversion of nucleons
to strange quark matter, $F$, is determined by
\begin{align}
\rho\frac{dF}{dy}=-\epsilon_\nu, \quad F=-\rho K\frac{dT}{dy}
\end{align}
where $\rho$ is the density, $K$ is the thermal conductivity of
strange quark matter, assumed to lie the range $10^{19}-10^{22} \mbox{
  cgs}$, and $y$ is the column density.  Again we will neglect slow
dependencies of $K$ on density and temperature so assuming that
$\rho$, $Q_\nu$ and $K$ are constant in the core this gives
\begin{align}
T=a(y+b)^{(-2/7)}\quad
a=\left(\frac{7}{2}\;\frac{7}{9}\frac{1}{(10^9)^8}\frac{Q_\nu}{\rho^2 K}\right)^{-1/7}
\end{align}
where $b$ is a constant of integration determined by the requirement
that all heat entering the core is emitted as
neutrinos, $\dot{m}Q_\mathrm{SQM}=\rho^{-1}\int \epsilon_\nu dy$. Assuming
that the integral is dominated by the surface contribution this gives
for the core surface temperature
\begin{align}\label{niso}
T_\mathrm{i,9}=0.87\, Q_{\nu,21}^{-1/9} K_{20}^{-1/9}\left(\frac{Q_\mathrm{SQM}}{100\mbox{ MeV}}\right)^{2/9}\left(\frac{\dot{m}}{0.3\;\dot{m}_\mathrm{Edd}}\right)^{2/9}
\end{align}
where $Q_{\nu,21}=Q_\nu/10^{21}$,
$K_{20}=K/10^{20}\mbox{ cgs}$ and $\dot{m}_\mathrm{Edd}\simeq
1.6\times10^{-8}\mbox{ M}_\odot\mbox{ yr}^{-1}/4\pi R^2m_u$ 
is the Eddington accretion rate in units of nucleons cm$^{-2}$ s$^{-1}$
with $R$ being the stars radius and $m_u$ the atomic mass unit. 

If however the thermal conductivity of strange quark matter is sufficiently
large the core becomes isothermal with the temperature given by $4\pi
R^2\dot{m}Q_\mathrm{SQM}=(4\pi R^3/3)Q_\nu T_\mathrm{i,9}^8$ so 
\begin{align}\label{iso}
T_\mathrm{i,9}=0.54\, Q_{\nu,21}^{-1/8}\left(\frac{Q_\mathrm{SQM}}{100\mathrm{
    MeV}}\right)^{1/8}\left(\frac{\dot{m}}{0.3\;\dot{m}_\mathrm{Edd}}\right)^{1/8}R_6^{-1/8}  
\end{align}
where $R_6$ is the core radius in units of 10 km. Equating this to
Eq.~\eqref{niso} shows that the core becomes isothermal at a critical
thermal conductivity $K_\mathrm{crit}=4.3\times 10^{21}\mbox{ cgs }
Q_{\nu,21}^{1/8}(Q_\mathrm{SQM}/100\mathrm{
  MeV})^{7/8}(\dot{m}/0.3\;\dot{m}_\mathrm{Edd})^{7/8}$. 

It should be noted that although Eqs.~\eqref{niso} and~\eqref{iso} are
certainly oversimplified they were shown by \cite{Page:2005} to agree
reasonably well with numerical calculations of a full general
relativistic model. Furthermore they offer a physical transparency,
which given the many poorly constrained properties of strange quark
matter, is perhaps just as desirable in what follows as full numerical
models. Specifically Eqs.~\eqref{niso} and~\eqref{iso} are general enough
not to depend on the properties of a particular strange quark matter
phase -- except for the assumption of slow neutrino emissivity
necessary to obtain superburst ignition. They are in fact general
enough to also describe neutron stars at the same level of detail, and
indeed Eq.~\eqref{iso} corresponds entirely to case (ii.a) in
\cite{Yakovlev:2003} for low-mass neutron stars with slow neutrino
emission. Thus much of what follows would also apply for neutron stars --
although in that case it has of course been done in much greater detail. 

In deriving Eqs.~\eqref{niso} and~\eqref{iso} we assumed that cooling by
neutrino emission dominated photon emission from the surface. This is
certainly the case in superbursters, where the core is very hot, but
for cold enough cores this may not be the case and one should take the
surface luminosity into account. Approximating the surface with a
blackbody we can estimate when this becomes important by setting,
$4\pi R^2 \dot{m}Q_\mathrm{SQM}=4\pi R^2\sigma T_\mathrm{S}^4$, which using
the \cite{Gudmundsson:1983} relation, $T_\mathrm{i,9}\simeq 0.1
T_\mathrm{S,6}^2$, gives a core temperature
\begin{align}
T_\mathrm{i,9}=20
\left(\frac{\dot{m}}{0.3\;\dot{m}_\mathrm{Edd}}\right)^{1/2}\left(\frac{Q_\mathrm{SQM}}{100\mbox{
    MeV}}\right)^{1/2}
\end{align}    
Equating this to Eq.~\eqref{iso} shows that cooling by photon emission
becomes important in the isothermal case below an accretion rate of 
\begin{align}
\dot{m}=3.1\times 10^{-13} Q_{\nu,21}^{-1/3}
\left(\frac{Q_\mathrm{SQM}}{100\mbox{ MeV}}\right)^{-1} R_6^{-1/3}
\mbox{ M}_\odot\mbox{ yr}^{-1}
\end{align}
with a similar expression giving a somewhat lower limit for a
non-isothermal core. Similar conclusions for neutron stars were
reached by \cite{Yakovlev:2003}. Given the right combination of core
parameters photon emission may therefore be important in soft X-ray
transients with the lowest accretion rates as low as a few times
$10^{-12} \mbox{ M}_\odot\mbox{ yr}^{-1}$. 

In the following subsections we shall ignore photon cooling from the
surface, which allows analytical constraints on the core parameters
and simple scaling relations for the surface temperature independent
of assumptions about the core parameters. This approach is justified
for high $Q_\mathrm{SQM}$, but we return to discuss the importance of
photon cooling in Sect.~\ref{sec4} as we shall find neutrino cooling unable to
account for the coldest soft X-ray transients.

\subsection{Constraints on the core parameters}
Eqs.~\eqref{niso} and~\eqref{iso} relate quark matter properties
represented by $Q_\nu$, $Q_\mathrm{SQM}$ and $K$ to the stellar
accretion rate and core surface temperature under the assumption that
the accretion rate is high enough for neutrino cooling to dominate.
In case there are independent constraints on the core surface
temperature and the accretion rate Eqs.~\eqref{niso} and~\eqref{iso} can
also be regarded as necessary relations between the quark matter core parameters
$Q_\nu$, $Q_\mathrm{SQM}$ and $K$. This approach was taken for superbursts in
\cite{Page:2005} assuming that the crust below the superburst ignition
depth is approximately isothermal so the core surface temperature can
be approximated by the superburst ignition temperature,
$T_\mathrm{i,9}\simeq 0.7$ at $\dot{m}=0.1-0.3\,
\dot{m}_\mathrm{Edd}$. We will make the same assumption and take over
the superburst parameter constraints from \cite{Page:2005}.

Assuming that soft X-ray transients are
also powered by deep crustal heating, a
different independent constraint on the quark core parameters
can be obtained by applying Eqs.~\eqref{niso} and~\eqref{iso}
to observations of the accretion rate and surface temperature 
of soft X-ray transients using a
$T_\mathrm{S}-T_{i}$ relation such as the one in \cite{Yakovlev:2004}
or \cite{Gudmundsson:1983}. 

\begin{figure}[!htbp]
\centering
\includegraphics[width=8.5cm]{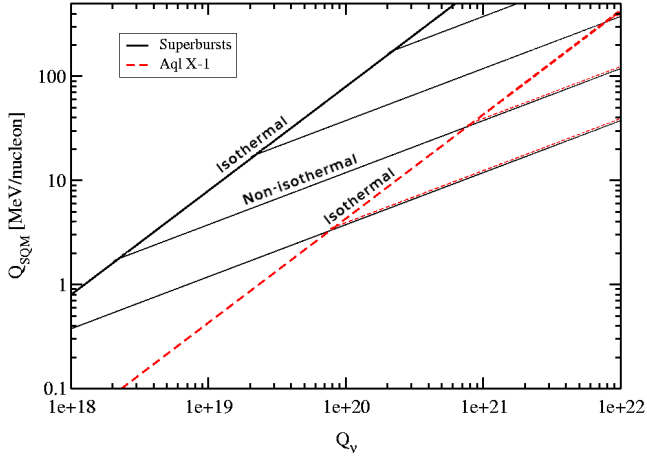}
\caption{Constraints on quark core parameters obtained from
  superbursters (solid)
  and the soft X-ray transient \object{Aql X-1} (dashed) assuming slow neutrino
  cooling for both. $Q_\mathrm{SQM}$ is given as a
  function of $Q_\nu$ for different choices of $K$. Thin lines show
  non-isothermal constraints from Eq.~\eqref{niso} with $K=10^{19},\,
  10^{20},\, 10^{21},\, 10^{22}\mbox{ cgs}$ from the bottom
  up. Non-isothermal constraints for \object{Aql X-1} are nearly
  indistinguishable from those for superbursters. Thick lines show
  isothermal constraints from Eq.~\eqref{iso}. We take $\dot{m}=0.3
  \dot{m}_\mathrm{Edd}$ for the superbursters. }
\label{fig1}
\end{figure}

\object{Aql X-1} is an example of a well studied
soft X-ray transient which has $\langle\dot{m}\rangle\simeq 10^{-10}
\mbox{ M}_\odot \mbox{ yr}^{-1}$, $T^\infty_S\simeq 113\mbox{ eV}$
\citep{Yakovlev:2003,Rutledge:2002} and therefore $T_\mathrm{i,9}\simeq 0.3$ using a
canonical mass-radius relationship, $R=10\mbox{ km}$, $M/R=0.2$, and
$T_\mathrm{S}=T_\mathrm{S}^\infty/(1-2GM/R)^{1/2}$. 
The relations between the quark matter core parameters 
thus obtained for superbursts and for the
soft X-ray transient \object{Aql X-1} are compared in
Fig.~\ref{fig1}. If we keep $Q_\nu$ fixed and increase
$Q_\mathrm{SQM}$ then $K$ must also increase to produce the required
temperature until the critical conductivity is reached and the star
must be isothermal. Therefore only the regions below the isothermal curves
contain possible solutions for each class of object 
and there can be no common solution for superbursters and
\object{Aql X-1} between the two isothermal curves shown. 
It is clear that the range of quark matter parameters permitted for 
the soft X-ray transient \object{Aql X-1} is significantly smaller 
than that allowed by the superbursters; in particular the quark matter
binding energy per baryon, $Q_\mathrm{SQM}$, necessary to fit the soft
X-ray transient must be in the lower part of the range allowed for fitting 
superbursters, other parameters being equal.
We further note that the
non-isothermal relations from Eq.~\eqref{niso} agree remarkably well
in the range where common parameters fit superbursters and 
\object{Aql X-1}, whereas no simultaneous isothermal solutions for the
two classes of objects seem to exist.

For a fixed (albeit unknown) set of strong interaction parameters
neither the quark matter binding energy $Q_\mathrm{SQM}$, nor the
quark core surface density should vary between stars, because both are
given directly by the strong interaction parameters.
The neutrino luminosity is dominated by the core surface contribution 
and strange stars are furthermore characterized by an almost constant
density in the core unless we are very close to the maximum mass strange
star, so the slow
variations with density of $Q_\nu$ and $K$ should be unimportant. 
$K$ may be inversely proportional to the
temperature \citep{Haensel:1991} and can therefore vary by a small 
factor between the coldest soft X-ray transients with quark core temperature
$\sim 10^8$ K and superbursters with core temperature $\sim 7\times 10^8$ K.
If no density thresholds between phases or emission
mechanisms are crossed when going from superbursters to soft X-ray
transients it
should therefore be reasonable to expect approximate agreement between
the relations in Fig.~\ref{fig1}, as seen for the non-isothermal
solutions, but not order of magnitude
differences as found in the isothermal case.  

\subsection{Scaling relations for the surface temperature}
In this subsection we will use Eqs.~\eqref{niso} and~\eqref{iso} to
relate the parameters for strange stars with
deep crustal heating that can fit soft X-ray transient data
and superbursters simultaneously, assuming slow neutrino cooling to be
the dominant cooling mechanism. In particular we will derive relations
between surface temperature and accretion rate for soft X-ray transients
for strange star model parameters that fit superbursts. These relations
will then be compared to observational data for soft X-ray transients
in the following Section.

To the extent that
$Q_\mathrm{SQM}$ and $Q_\nu$ do not vary between stars we can utilize 
Eqs.~\eqref{niso} and~\eqref{iso} to
compare the other parameters for two arbitrary strange stars with
deep crustal heating. In the following we shall in particular
be interested in cases where one of the stars is a superburster (SB) and the
other a soft X-ray transient (X).
From Eq.~\eqref{niso} we have in the non-isothermal case
\begin{align}\label{ABniso}
\begin{split}
&\alpha_\mathrm{SB}\equiv\left(T_\mathrm{i,9}^{9/2}\left[\frac{\langle\dot{m}\rangle}{0.3\,\dot{m}_\mathrm{Edd}}\right]^{-1}\right)_\mathrm{SB}=\\&\left(\frac{K_\mathrm{20,X}}{K_\mathrm{20,SB}}\right)^{1/2}\left(T_\mathrm{i,9}^{9/2}\left[\frac{\langle\dot{m}\rangle}{0.3\,\dot{m}_\mathrm{Edd}}\right]^{-1}\right)_\mathrm{X}
\end{split}
\end{align}
and from Eq.~\eqref{iso} in the isothermal case
\begin{align}\label{ABiso}
\begin{split}
&\alpha_\mathrm{SB}\equiv\left(T_\mathrm{i,9}^{8}\left[\frac{\langle\dot{m}\rangle}{0.3\,\dot{m}_\mathrm{Edd}}\right]^{-1}\right)_\mathrm{SB}=\\&\left(\frac{R_{6,X}}{R_{6,SB}}\right)\left(T_{9,i}^{8}\left[\frac{\langle\dot{m}\rangle}{0.3\,\dot{m}_\mathrm{Edd}}\right]^{-1}\right)_\mathrm{X} .
\end{split}\end{align}

In a soft X-ray transient the temperature at the quark core surface can be
obtained from a $T_\mathrm{S}-T_{i}$ relation with
$T_\mathrm{S}=T_\mathrm{S}^\infty/(1-2GM/R)^{1/2}$. 
The average accretion rate can be
estimated from the outburst luminosity, $L_o\sim (\Delta M/t_o)(GM/R)$,
where $\Delta M$ is the accreted mass and $t_o$ is the outburst
duration, so if $t_q$ is the time spent in quiescense
$\langle\dot{m}\rangle= \frac{\Delta M/m_u}{4\pi R^2(t_o+t_q)}$.
In Eqs.~\eqref{ABniso} and~\eqref{ABiso} we should therefore further allow
for different $M/R$ ratios. If we use a detailed $T_\mathrm{S}-T_{i}$
relation such as in \cite{Yakovlev:2004} there is no point in doing
this analytically. As noted such relations can however be approximated
reasonably well by $T_{i,9}\simeq
0.1\,T_\mathrm{S,6}^2$ which allows a physically transparent
interpretation. 

We will use Eqs.~\eqref{ABniso} and~\eqref{ABiso} to
compare superbursters and soft X-ray transients so writing
$\alpha_\mathrm{SB}$ for the left hand side scaled to fit the superburst
data and dropping the subscript on the soft X-ray transient parameters we get
\begin{align}
\begin{split}T_\mathrm{S,6}=&10^{1/2}\left[\alpha_\mathrm{SB} \left(\frac{K_\mathrm{20,SB}}{K_\mathrm{20}}\right)^{1/2}\frac{\langle\dot{m}\rangle}{0.3\,\dot{m}_\mathrm{Edd}}\right]^{1/9}\label{nisotest}\end{split}\\
\alpha_\mathrm{SB}=&0.2\left(\left(\frac{T_\mathrm{i,9}}{0.7}\right)^{9/2}\left[\frac{\langle\dot{m}\rangle}{0.3\,\dot{m}_\mathrm{Edd}}\right]^{-1}\right)_\mathrm{SB}
\end{align} 
in the non-isothermal case and correspondingly in the isothermal case
\begin{align}
\begin{split}T_\mathrm{S,6}=&10^{1/2}\left[\alpha_\mathrm{SB}
    \left(\frac{R_\mathrm{6,SB}}{R_6}\right)\frac{\langle\dot{m}\rangle}{0.3\,\dot{m}_\mathrm{Edd}}\right]^{1/16}\label{isotest}\end{split}\\
\alpha_\mathrm{SB}=&0.058\left(\left(\frac{T_\mathrm{i,9}}{0.7}\right)^{8}\left[\frac{\langle\dot{m}\rangle}{0.3\,\dot{m}_\mathrm{Edd}}\right]^{-1}\right)_\mathrm{SB}\;
.
\end{align} 
As a last possibility we note from Fig.~\ref{fig1} that solutions can
be found with non-isothermal superbursters and isothermal soft X-ray
transients. In this case we use $K_\mathrm{crit}$ in Eq.~\eqref{niso} to
obtain a relation between $Q_\mathrm{SQM}$ and $Q_\nu$ which when
inserted in Eq.~\eqref{iso} gives
\begin{align}
\begin{split}T_\mathrm{S,6}=&10^{1/2}\left[\alpha_\mathrm{SB} \frac{\langle\dot{m}\rangle}{0.3\,\dot{m}_\mathrm{Edd}}\right]^{1/9}\label{mixed}\end{split}\\
\alpha_\mathrm{SB}=&0.154\left(\left(\frac{T_\mathrm{i,9}}{0.7}\right)^{9/2}\left[\frac{\langle\dot{m}\rangle}{0.3\,\dot{m}_\mathrm{Edd}}\right]^{-1}\right)_\mathrm{SB}
\end{align} 

These expressions can be used to test the constancy of the quark core
parameters when going from superbursters to soft X-ray transients, 
with $\alpha_\mathrm{SB}$ incorporating the constraints
imposed by superbursters. We note that small variations in the
quantities in square brackets will be strongly suppressed and
specifically it will not matter whether we use a canonical mass radius
relation for neutron stars to find the accretion rate or one more
relevant to strange stars where the radius may be smaller by a few
kilometers. This would translate into a temperature correction of
about one percent which is far below the observational uncertainty so
we can use the accretion rates in the literature even though they are
derived for neutron stars. There is also no reason to expect that
$(R_\mathrm{6,SB}/R_6)^{1/16}$ will deviate much from unity so we drop this
in the following. The $(K_\mathrm{SB}/K)^{1/18}$ term can give a
factor $\sim 0.9$ if $K$ depends on the temperature and
$\alpha_\mathrm{SB}\lesssim 0.2$ so we keep both of these terms but
note that uncertainties in these parameters are strongly suppressed too.

\section{Comparison with observations}\label{sec4}
\begin{table*}[!htbp]
\centering \footnotesize
\begin{tabular}{|l|c|c|c|c|}
\hline
Source & $kT_\mathrm{S}$ [eV] & $R$ & $\langle\dot{m}\rangle$ $[M_\odot\mathrm{
    yr}^{-1}]$ & Reference\\
\hline
\object{Aql X-1}         &$113^{+3}_{-4}$&13.2&$10^{-10}$&\cite{Rutledge:2002,Yakovlev:2003}\\
\object{Cen X-4}         &76 $\pm 7$, 85$\pm3$&12.9&$1.9\times 10^{-11}$&\cite{Rutledge:2001,Campana:2004,Tomsick:2004}\\
\object{4U 1608-522}     &170$\pm 30$&9.4 &$6.9\times 10^{-11}$&\cite{Rutledge:1999,Tomsick:2004}\\
\object{SAX J1808.4-3658}&$<54$      &--  &$5\times 10^{-12}$  &\cite{Campana:2002,Yakovlev:2004b}\\
\object{SAX J1810.8-2609}&$<72$      &--  &$1.7\times 10^{-12}$&\cite{Jonker:2004,Tomsick:2004}\\
\object{XTE J2123-058}   &$<66$      &--  &$2.9\times 10^{-12}$&\cite{Tomsick:2004}\\
\object{XTE J1709-267}   &$116 \pm 2$&--  &$1.2\times 10^{-10}$&\cite{Jonker:2003,Jonker:2004}\\
\hline
\end{tabular}
\caption{Properties of the observational sample commonly used to test
  deep crustal heating. The temperature quoted for \object{Aql X-1} is the
  redshifted $T_\mathrm{S}^\infty$. We use $T_\mathrm{S}=T_\mathrm{S}^\infty/(1-2GM/R)^{1/2}$ with the
  canonical $M/R=0.2$. For \object{Cen X-4} we quote the temperatures from both
  \cite{Rutledge:2001} and \cite{Campana:2004}.}
\label{tab1}
\end{table*} 
\normalsize

\begin{figure*}[!ht]
\centering
\includegraphics[width=17cm]{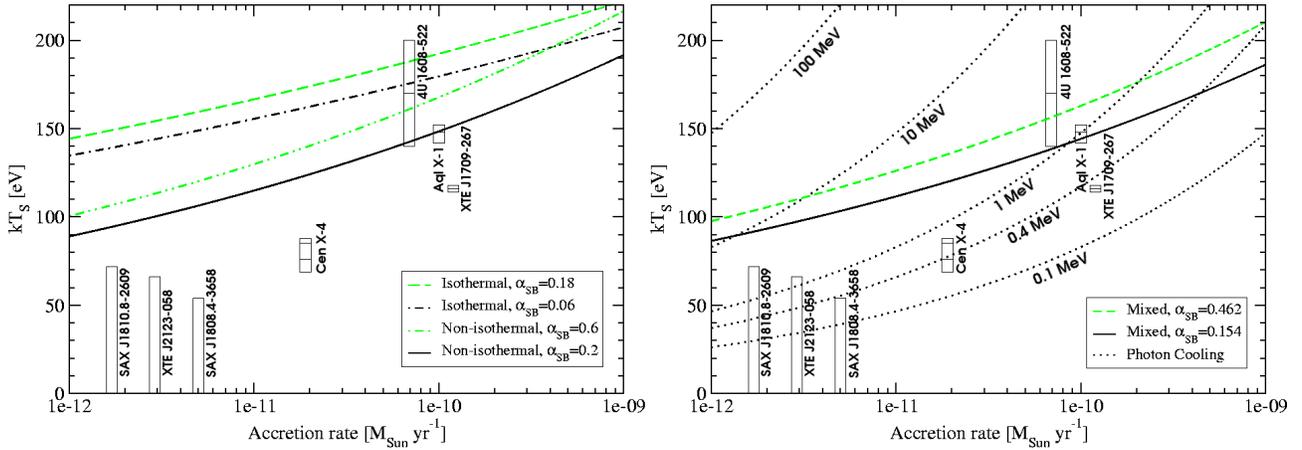}
\caption{{\it Left}: comparing cooling by a non-isothermal (dashed and
  dot-dashed; cf. Eq.~\eqref{nisotest}) and isothermal (dashed-doubledot
  and solid; cf. Eq.~\eqref{isotest})
  core to the observations in
  Table~\ref{tab1}. For \object{Cen X-4} the quoted temperature intervals
  overlap. {\it Right}: the ``mixed'' case with non-isothermal superbursts
  and isothermal soft X-ray transients (dashed and solid; cf. Eq.~\eqref{mixed}) and
  photon cooling (dotted) assuming a
  pure blackbody with
  $Q_\mathrm{SQM}$ as labelled for each curve.}
\label{fig2}
\end{figure*}

The observational sample most commonly used to test relations for deep
crustal heating such as Eqs.~\eqref{isotest} and~\eqref{nisotest}
consists of soft X-ray transients in the field with short outburst
durations and recurrence times, for which X-ray observations in
quiescense have determined or limited the surface temperature. The
outburst duration should be short so the crust has not been
significantly heated by the release of nuclear energy
from the accreted matter and the surface temperature is therefore set
by deep crustal heating. The recurrence time should be short enough
that at least a few outbursts have been observed allowing a reasonable
estimate of the time averaged accretion rate -- although one should
remember that the last 37 years of X-ray observations are not
necessarily representative of the $\sim 10^4$ years of accretion
history needed to reach thermal equilibrium. Similarly since the
accretion history for stars in globular clusters is uncertain one
should only use systems in the field.

Useful lists of such systems and their properties including
references can be found in e.g. \cite{Yakovlev:2003} and \cite{Tomsick:2004}. We
will take this data, summarized in Table~\ref{tab1}, as our
observational basis and briefly discuss each system below.
The best examples of such systems are \object{Aql X-1}, \object{Cen X-4} and \object{4U 1608-522}
being short outburst field soft X-ray transients with detected thermal
components. \cite{Tomsick:2004} discuss the outburst history of these
sources and we cite their accretion rates which are based on the
review of soft X-ray transient activity by \cite{Chen:1997} using
\begin{align}\label{mdot}
\langle\dot{m}\rangle&=\frac{\bar{L}_\mathrm{Peak}Nt_\mathrm{o}f}{\epsilon
  c^2 (33\mbox{ yr})}
\end{align}
where $\bar{L}_\mathrm{Peak}$ is the time averaged peak outburst
luminosity, $N$ is the number of outbursts, $f=0.4$ is a factor which
accounts for the shape of the light curve, $\epsilon=0.2$ is the
fraction of rest mass energy released during the accretion and the
time average is taken over 33 years of observation. \object{Aql X-1}, \object{Cen X-4}
and \object{4U 1608-522} have gone though 21, 2 and 16 outburst respectively
between 1969 and 2003 allowing estimates of the accretion
rate. Temperatures are found from spectral fits to absorbed neutron star
atmosphere models assuming a mass of $1.4 \mbox{ M}_\odot$ -- this
also yields the best fit radius cited. \object{Cen X-4} and some observations
of \object{Aql X-1} further require a power law component possibly indicating
residual accretion during quiescense.

For \object{SAX J1808.4-3658}, \object{SAX J1810.8-2609} and \object{XTE J2123-058} no thermal
components have been detected and so we only cite upper limits to the
surface temperature. \object{SAX J1810.8-2609} and \object{XTE J2123-058} have each
shown only one outburst so the accretion rate is actually an
upper limit. The accretion rate cited for \object{SAX J1808.4-3658} is based on
the two outbursts in 1996 and 1998, but further outburst have ocurred
in 2000, 2002 and 2005 (see \cite{Wijnands:2004a} for a review). These
systems are therefore not expected to agree well with the theory, but
at least the upper limits to their surface temperatures should not lie
significantly below the prediction of deep crustal heating if they
have accreted for long enough to have reached thermal equilibrium.

\object{XTE J1709-267} lies 9-10 arcmin from NGC 6293 which has a
tidal radius of 14.2 arcmin and so may be associated with this
globular cluster although distance measurements are inconclusive on
this \citep{Jonker:2004}. It has shown two outbursts since 1995 and
has a varying power law component.

As should be clear from the discussion above the data is rather
uncertain and should be treated with some caution. In particular the
time averaged accretion rates are hardly more than estimates, and the
surface temperatures, when they can be found, are uncertain and may
not only reflect deep crustal heating.  Nevertheless deep crustal
heating should provide a minimum temperature for systems that have
accreted long enough and so we should at least test for consistency
with the observations if not perfect agreement. In Fig.~\ref{fig2} we
plot the data from Table~\ref{tab1} and compare with photon cooling
curves as well as Eqs.~\eqref{nisotest}, ~\eqref{isotest}
and~\eqref{mixed}, that assume neutrino cooling, 
for values of $\alpha_\mathrm{SB}$
corresponding to $\dot{m}_\mathrm{SB}=0.1$ and $0.3 \,\dot{m}_\mathrm{Edd}$.

The overall agreement with our predictions from neutrino emission in
superbursts is not impressive. Except \object{Aql X-1} and \object{4U
1608-522} the observations fall significantly below the predicted
temperature for neutrino cooling, and actually pure photon 
cooling does better if
$Q_\mathrm{SQM}\lesssim 1$ MeV although it should not be dominant for
the high accretion rates in \object{Aql X-1} and \object{4U 1608-522}. This
raises two issues.

First it is noteworthy
that the two sources which fall close to the superburst prediction
without photon cooling,
are the hottest and most frequently accreting sources thus resembling
superbursters the most. In turn this raises the question whether the
accretion history may lead to heating with a pronounced time
dependence and whether the time averaged accretion rate is then still
a good predictor for the temperature -- we discuss this possibility in
the next section.
 
\begin{figure*}[!ht]
\centering
\includegraphics[width=17cm]{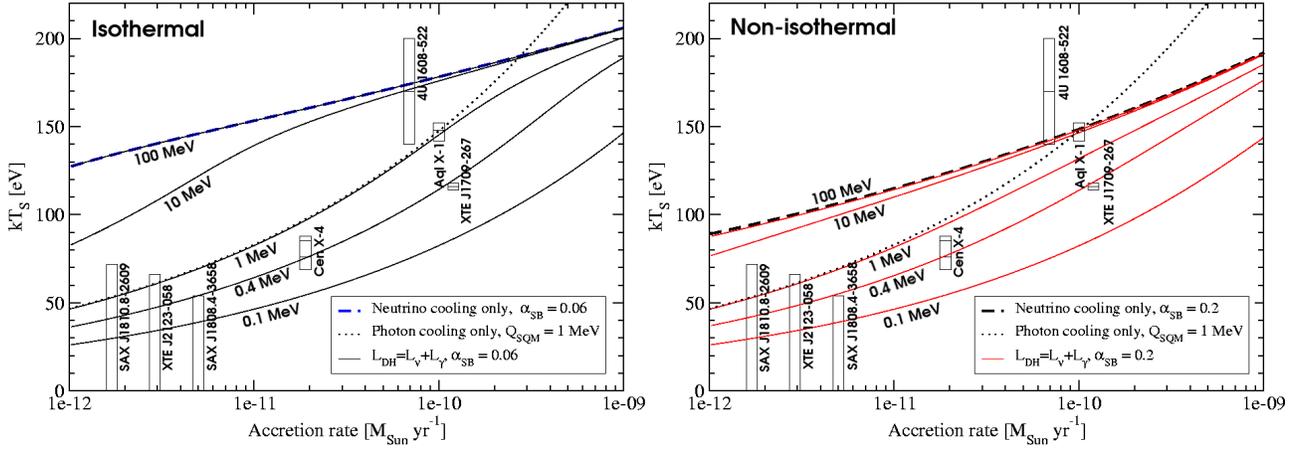}
\caption{{\it Left}: comparing combined photon and neutrino cooling of
  an isothermal core (solid; cf. Eq.~\eqref{isocombined}) to the
  observations in Table~\ref{tab1} with $Q_\mathrm{SQM}$ as labelled
  for each curve . For reference we also show curves with neutrino
  cooling only (dashed; cf. Eq.~\eqref{isotest}) and photon cooling
  only (dotted). The curves assuming $Q_\mathrm{SQM}=100$ MeV and
  neutrino cooling only are nearly identical. For \object{Cen X-4} the
  quoted temperature intervals overlap. {\it Right}: corresponding
  curves for a non-isothermal core.}
\label{fig3}
\end{figure*}

Second we see that if $Q_\mathrm{SQM}$ is sufficiently low, photon
emission from the surface may be very important and so we should
combine the two cooling mechanisms. Thus we now demand -- similarly to
the procedure in \cite{Yakovlev:2003} -- $L_\mathrm{DH}=4\pi
R^2\dot{m}Q_\mathrm{SQM}=L_\gamma+L_\nu$ where $L_\mathrm{DH}$ is the
luminosity required to carry away the energy released by deep crustal
heating and $L_\gamma=4\pi R^2\sigma T_\mathrm{S}^4$ is the usual
blackbody luminosity. Following the calculation that leads to
Eq.~\eqref{niso} the non-isothermal neutrino luminosity is
\begin{align}
L_\nu=4\pi R^2\cdot0.3\, \dot{m}_\mathrm{Edd}\cdot 100 \mbox{ MeV}\cdot
\left(\frac{T_\mathrm{i,9}}{0.87}\right)^{9/2} Q_{\nu,21}^{1/2} K_{20}^{1/2}\;.\label{Lnu1}
\end{align} 
Neutrino cooling should dominate in superbursters,
$L_\mathrm{DH}=L_\nu$ with $T_\mathrm{i,9}^\mathrm{SB}=0.7$, so using the
constraints this imposes on the core parameters  Eq.~\eqref{Lnu1} gives
\begin{align}
L_\nu=4\pi
R^2\left(\frac{T_\mathrm{i,9}}{0.7}\right)^{9/2}\,\dot{m}_\mathrm{SB}\,Q_\mathrm{SQM}\;,\label{Lnu}
\end{align}
and using $T_\mathrm{i,9}=0.1\,T_\mathrm{S,6}^2$ the condition
$L_\mathrm{DH}=4\pi R^2\dot{m}Q_\mathrm{SQM}=L_\gamma+L_\nu$ then becomes
\begin{align}
\begin{split}\dot{m}\,Q_\mathrm{SQM}=&\left(\frac{0.1}{0.7}\right)^{9/2}
  \dot{m}_\mathrm{SB}\,Q_\mathrm{SQM} T_\mathrm{S,6}^9+\sigma T_\mathrm{S}^4\;.\end{split}\label{combined}
\end{align}
Correspondingly in the isothermal case 
\begin{align}
\begin{split}\dot{m}\,Q_\mathrm{SQM}=&1.8\times 10^{-7}
  \dot{m}_\mathrm{SB}\,Q_\mathrm{SQM} T_\mathrm{S,6}^{16}+\sigma T_\mathrm{S}^4\;.\end{split}\label{isocombined}
\end{align}
This can be solved numerically and in Fig.~\ref{fig3} we show
representative solutions for a range in $Q_\mathrm{SQM}$ and compare
with pure photon and neutrino cooling. We see that photon
cooling dominates at low accretion rates and low $Q_\mathrm{SQM}$
whereas for high accretion rates and $Q_\mathrm{SQM}$ we recover the
neutrino cooling curves relevant to superbursters. In particular we
note that $Q_\mathrm{SQM}=0.4\mbox{ MeV}$ while still fulfilling the
superburst requirements at high accretion rates provides a close fit
to some of the sources at low accretion rates that were difficult to
explain with neutrino cooling alone. In this case other processes must
be heating the surfaces of \object{Aql X-1} and \object{4U 1608-522}
though. 

\cite{Brown:1998} similarly found that deep crustal heating of
neutron stars could explain the quiescent emission from \object{Cen
  X-4} if only 0.1 MeV of the energy released by pycnonuclear
reactions during an outburst was conducted into the
core and deposited there.~\cite{Yakovlev:2003,Yakovlev:2004}~instead interpreted \object{Cen X-4} and
\object{SAX J1808.4-3658} as massive neutron stars with enhanced
neutrino emission. In the context of strange stars the energy is
deposited directly in the core, but no enhanced neutrino cooling is
required if the quark matter binding energy $Q_\mathrm{SQM}$ is small.

\section{Time dependence of the mass transfer rate}\label{sec5}
The quiescent luminosity from some of the sources discussed here
 (i.e.~Aql X-1~\citep{Rutledge:2002}, and Cen
 X-4~\citep{Paradijs:1987, Campana:1997,Campana:2004,Rutledge:2001})
 have been found to vary by factors of a few on timescales of days to
 years raising the question of how time dependent the heating
 mechanism is following an outburst. Furthermore the time dependence
 of the heating mechanism may be important in sources with long
 recurrence times and help explain their relatively low temperatures
 reducing the need to keep $Q_\mathrm{SQM}$ small.

To fully demonstrate either of these points would require a detailed
time dependent model for cooling strange stars, which is beyond our
scope here. Thus the calculations below are concerned only with the
time dependence of the tunneling rate -- and hence the heating --
following an outburst, but we note that with respect to its
consequences for the surface temperature we may be guided by similar
considerations for neutron stars. 

In neutron stars heat is released during and after an accretion event
at the depth and on the time scale (months for the pycnonuclear
reactions) corresponding to the individual electron captures, neutron
emissions and pycnonuclear reactions. It then takes about 1 yr for the
heat from the pycnonuclear reactions to diffuse to the surface
\citep{Brown:1998,Ushomirsky:2001} and since it spreads out along the
way, it is seen there as a heat wave lasting approximately the thermal
diffusion time to the surface. \cite{Ushomirsky:2001} investigated the
time dependence of this process for 1-day long outbursts, and found
that energy deposited at depths, where the diffusion time to the
surface was shorter than the outburst recurrence time, could produce
significant variations in the surface luminosity following a pattern
in time corresponding to the diffusion time for each nuclear
reaction. If for instance the recurrence time was 1 year and assuming
a low thermal conductivity in the crust, the luminosity variations
were a few percent of the average, and only one heat wave could be
seen because the diffusion time from the depth of the pycnonuclear
reactions was on the order of the recurrence time. If the recurrence
time was 30 years the variations in the luminosity could be up to 30
percent for cores with standard cooling and a factor of 3-4 for cores
with accelerated cooling (e.g. pion condensates). In this case two
separate heat waves were seen -- one corresponding to electron
captures in the outer crust visible a few days after the outburst and
the other to the pycnonuclear reactions deeper in the crust emerging
about a year later. For higher thermal conductivity variations were
larger and happened on shorter time scales.

By analogy if the tunneling rate at the bottom of the crust -- and
hence the heating -- is sufficiently time dependent following an
accretion event, we would expect the qualitative conclusions in
\cite{Ushomirsky:2001} to apply, possibly leading to strong variations
in the surface temperature following an outburst as the resulting
heat waves reach the surface. One difference may
be though that in strange stars the heat is released at a lower
density in the crust with a shorter diffusion time to the surface so
the two heat waves may be closer in time.

\begin{figure}[!htbp]
\centering
\includegraphics[width=8.5cm]{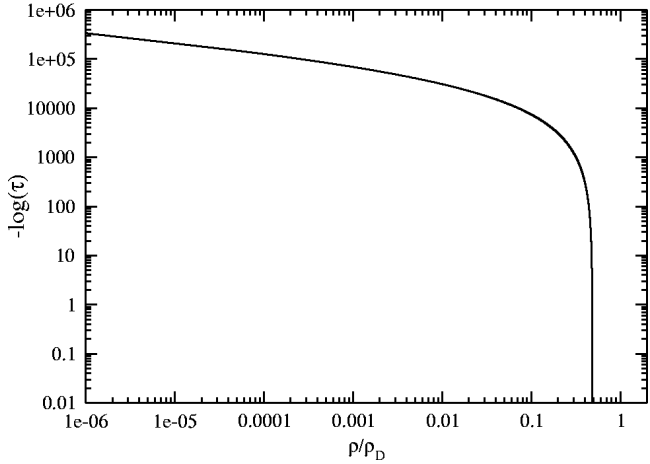}
\caption{Plot of Eq.~\eqref{tau} assuming $e\phi_e(R_S)=20$ MeV and
  \element[][106]{Ge} at the crust boundary. Note that for this barrier height
  the gap width goes to zero and $\tau$ to 1 somewhat below the
  neutron drip density, which can then not be reached.}
\label{fig4}
\end{figure}

The transfer of mass from crust to core by tunneling takes place at a
rate of \citep{Alcock:1986,Stejner:2005}
\begin{align}
\begin{split}\frac{\mathrm{d}M_\mathrm{C}}{\mathrm{d}t} &=-m_\mathrm{N}N_\mathrm{ion}\times \nu\times \tau(\rho_\mathrm{b})\\
&=-4\pi R^2d\,\rho_\mathrm{b}\times \nu\times \tau(\rho_\mathrm{b})\\
&=-23.8\frac{\rho_\mathrm{b}}{\rho_\mathrm{D}}\tau(\rho_\mathrm{b}) \mbox{ M}_\odot \mbox{ s}^{-1}\end{split}\label{dMdt}
\end{align}
where $m_\mathrm{N}$ is the mass of ions at the bottom of the crust, $N_\mathrm{ion}$
is the number of such ions whose motion allow them to strike the
Coulomb barrier taken to be the number within one lattice distance,
$d\sim 200$ fm, $\rho_\mathrm{b}$ is the density at the bottom of the crust,
$\nu\lesssim 1$ MeV is the frequency of this motion and $\tau(\rho)$ is
the transmission rate. In \cite{Stejner:2005} we found that the
transmission rate can be approximated by
\begin{align}
\begin{split}\log{\tau(\rho)}=&-51Z\sqrt{\frac{2}{\Delta\phi_\mathrm{g}}}\Bigg[\arccos{\left(\frac{m_\mathrm{N}\Delta\phi_\mathrm{g}}{Ze\phi_e(R_\mathrm{S})}\right)^{1/2}}-\\
&\left(\frac{m_\mathrm{N}\Delta\phi_\mathrm{g}}{Ze\phi_e(R_\mathrm{S})}-\left(\frac{m_\mathrm{N}\Delta\phi_\mathrm{g}}{Ze\phi_e(R_\mathrm{S})}\right)^2\right)^{1/2}\Bigg]\label{tau}\end{split}\\
\Delta\phi_\mathrm{g}=1.9&\times 10^{-4} \left(\frac{\rho_\mathrm{b}}{\rho_\mathrm{D}}\right)^{1/3}\left(\frac{56}{A}\right)^{4/3}\sqrt{Z^{8/3}-1}
\end{align}
where $Z$ is the ion charge number, $A$ is the mass number,
$e\phi_e(R_\mathrm{S})$ is the height of the Coulomb barrier (i.e. the electric
potential at the core surface, $R_\mathrm{S}$) and
$\Delta\phi_\mathrm{g}=\phi_\mathrm{g}(R)-\phi_\mathrm{g}(R_S)$ is the difference in the
(Newtonian) gravitational potential between the surface of the star
and the core. We plot this behaviour in Fig.~\ref{fig4} assuming
$e\phi_e(R_\mathrm{S})=20\mbox{ MeV}$ and the crust composition in
\cite{Haensel:2003} (\element[][106]{Ge} at maximum density). From
Eq.~\eqref{dMdt} a star accreting $10^{-10}\mbox{ M}_\odot \mbox{
  yr}^{-1}$ must have $\log{\tau} \lesssim -42$ in equilibrium, so as
previously mentioned the tunneling rate close to equilibrium is very
sensitive to the density.

\begin{figure}[!htbp]
\centering
\includegraphics[width=8.5cm]{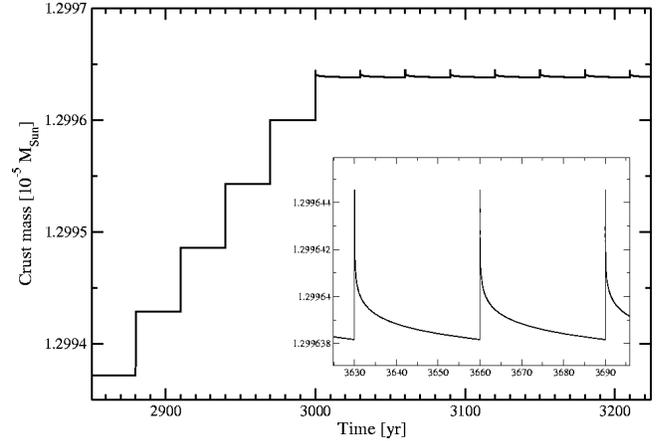}
\caption{A crude model of \object{Cen X-4}. Crust mass in a system accreting
with 10 day long accretion outbursts separated by quiescent intervals
of 30 years and an average accretion rate of $1.9\times 10^{-11}
\mbox{ M}_\odot \mathrm{ yr}^{-1}$. The inset shows a few equilibrium
cycles. The peaks seen here have a flat top which is not resolved
in time on the scale shown.}
\label{fig5}
\end{figure}

\begin{figure}[!htbp]
\centering
\includegraphics[width=8.5cm]{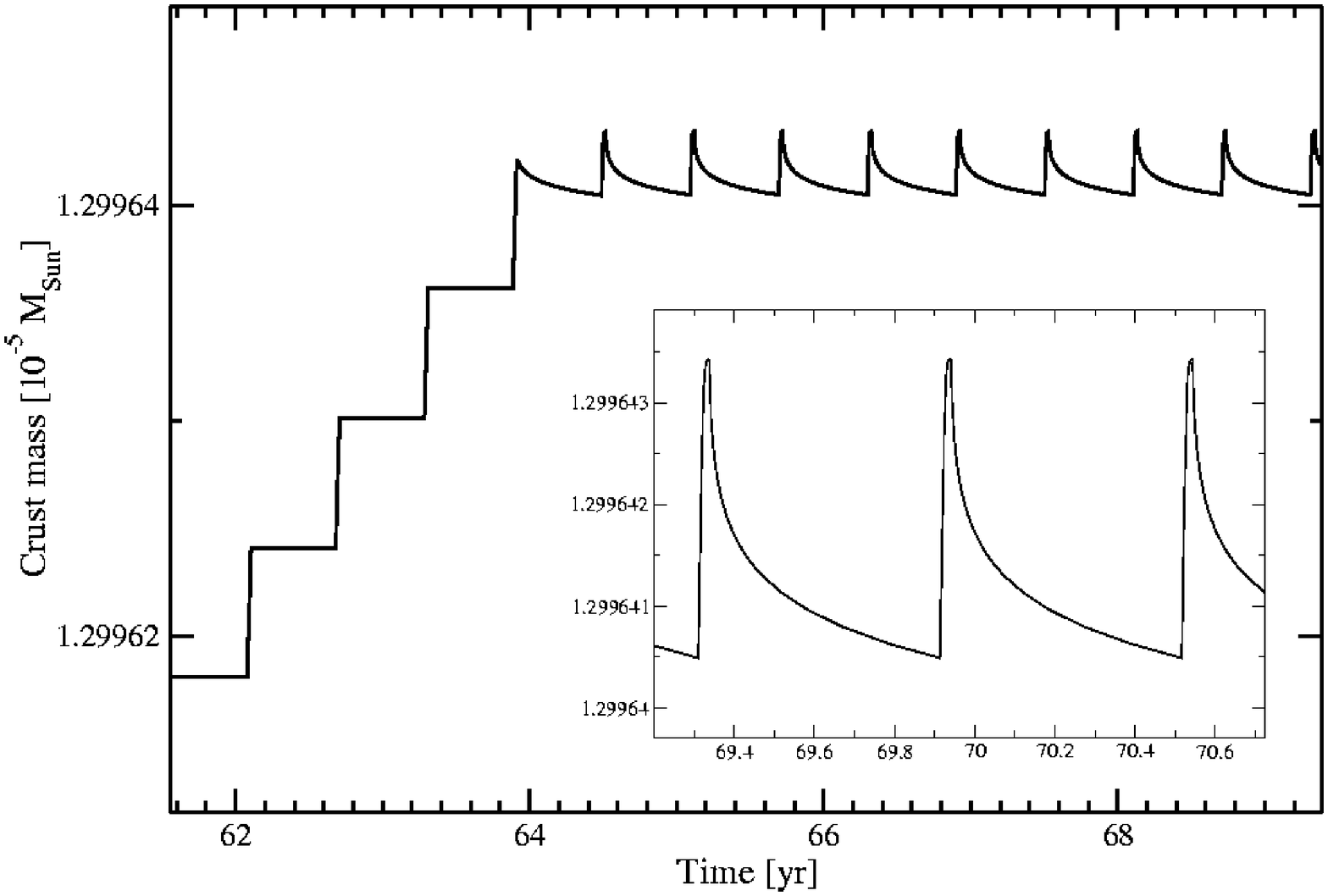}
\caption{A crude model of \object{Aql X-1}. Crust mass in a system accreting
with 10 day long accretion outbursts separated by quiescent intervals
of 220 days and an average accretion rate of $ 10^{-10}
\mbox{ M}_\odot \mathrm{yr}^{-1}$. The inset shows a few equilibrium
cycles.}
\label{fig6}
\end{figure}

Eq.~\eqref{mass} gives a relation between $M_\mathrm{C}$ and
$\rho_\mathrm{b}$ so we can solve Eq.~\eqref{dMdt} to find the crust
mass and tunneling rate for given accretion scenarios. Assuming the
same parameters as in Fig.~\ref{fig4} we show a solution in
Fig.~\ref{fig5} with 10 day long accretion outbursts separated by
quiescent intervals of 30 years and an average accretion rate of
$1.9\times 10^{-11} \mbox{ M}_\odot \mbox{ yr}^{-1}$ -- roughly
corresponding to \object{Cen X-4}. We start the integration with a
crust mass sufficiently low that the tunneling rate is negligible and
let it build up until the mass transferred to the core during each
cycle matches the average accretion rate. At this point the crust mass
is at its maximum -- as determined by the choice of $e\phi_e(R_S)= 20$
MeV -- during the accretion outbursts and then relaxes to a state with
very little tunneling within about a year after each
outburst. Fig.~\ref{fig6} shows the corresponding result for a system
resembling \object{Aql X-1} accreting 10 in every 220 days with an
average accretion rate of $10^{-10}\mbox{ M}_\odot \mbox{
yr}^{-1}$. Here we more clearly see the crust reach a maximum mass
during the 10 day long outbursts.

Thus practically all the heating takes place during and immediately
after the outburst when the transfer rate is essentially the accretion
rate, and the heating mechanism is indeed extremely time dependent. In
systems such as \object{Aql X-1} and \object{4U 1608-522} the outburst
recurrence time is comparable to the diffusion time to the surface
however. The heat wave resulting from an accretion outburst will
therefore be smeared out, and the crust temperature will not vary
significantly between outbursts unless one assumes a high thermal
conductivity in the crust, so a steady state model using the average
accretion rate should be able to predict the temperature. Conversely
in systems such as \object{Cen X-4} where the recurrence time is
significantly longer than the diffusion time to the surface one would
expect larger temperature variations following an outburst.

Furthermore in such systems the core -- or the surface layers of the
core in the non-isothermal case -- may have time to cool significantly
between outbursts. In this case we would not expect
Eqs.~\eqref{nisotest},~\eqref{isotest} and~\eqref{mixed} to hold, but
if (speculatively) the tunneling rate late in the cycle is a better
predictor of the temperature at this point, Eq.~\eqref{nisotest}
assuming neutrino cooling only would actually give a surface
temperature around 76 eV for \object{Cen X-4} as observed. The same
argument for \object{XTE J1709-267} assuming a 6 year recurrence time
would lead to surface temperatures as low as 83 eV.  Here the
recurrence time may not be quite long enough however, and furthermore
the temperature we cite was measured in May 2003 not very long after
the March-April 2002 outburst, so the argument may be too simple. The
temperature we cite for \object{Cen X-4} was measured in 1997-2000
long after the 1979 outburst so this should provide a more clear-cut
case. For the temperature of the three remaining sources to reach
their upper bounds in this way with the cited accretion rates, the
outburst recurrence times would have to be very long ($\sim$ 50, 100
and 800 years for \object{SAX J1810.8-2609}, \object{XTE J2123-058}
and \object{SAX J1808.4-3658} respectively). 

This shows how the 
the time dependence of the heating mechanism may potententially help
explain the coldest soft X-ray transients, even if $Q_\mathrm{SQM}$ is
too high for photon cooling to dominate. More work including
details on the cooling of the core and the transport properties of
both crust and core would be required though to show that it is
possible for the temperature to change this rapidly. 

\section{Conclusion}\label{sec6}
Using a simple scaling argument for the steady state temperature of
accreting strange stars we have presented a consistency check between
the constraints on strange quark matter derived from superburst
ignition conditions, and the temperature of quiescent soft X-ray
transients -- both assumed to derive from deep crustal heating by
transfer of matter from crust to core. 

For systems with short recurrence times where the steady state
temperature may be expected to be a good approximation we found
reasonable agreement assuming neutrino cooling only -- although the
range of quark matter parameters giving superbursters between the
isothermal curves in Fig.~\ref{fig1} is then excluded by soft X-ray
transients. Assuming a combination of photon and neutrino cooling for
soft X-ray transients, we
saw in Fig.~\ref{fig3} that if $Q_\mathrm{SQM}\lesssim 1$ MeV even
the coldest sources may be explained while still fulfilling superburst
ignition conditions at high accretion rates.

We further showed that the tunneling rate near the equilibrium crust
mass depends sensitively on the crust density and consequently that
essentially all the heating takes place during or immediately after an
outburst. This points to the need for a fully time dependent model to
fit systems with recurrence times significantly longer than the
thermal diffusion time to the surface, but if the low level tunneling
rate long after the outburst is a better predictor for such systems,
it would provide a fit for \object{Cen X-4} assuming only neutrino
cooling as well. The upper temperature bounds on \object{SAX
J1810.8-2609}, \object{XTE J2123-058} and \object{SAX J1808.4-3658}
can not be reached for the cited accretion rates in this way unless
the recurrence times are very long. This is consistent with the
outburst histories of \object{SAX J1810.8-2609} and \object{XTE
J2123-058} -- one known outburst each -- but does not compare well
with the $\sim 2$ yr recurrence time for \object{SAX
J1808.4-3658}. Similarly to the situation with neutron stars
\citep{Campana:2002,Yakovlev:2004} some form of enhanced cooling would
thus seem to be required if \object{SAX J1808.4-3658} has accreted for
long enough to heat the core to its equilibrium temperature and
$Q_\mathrm{SQM}$ is too high for photon cooling to
dominate. \object{XTE J1709-267} has an intermediate recurrence time,
and its temperature was measured shortly after an outburst making
interpretation difficult.

If -- as seems reasonable -- a combination of the two cooling
mechanisms is required to explain the observations we should further
note that the constraints imposed by \object{Aql X-1} in
Fig.~\ref{fig1} are not relevant since they were derived assuming only
neutrino cooling. The requirement that $Q_\mathrm{SQM}\lesssim 1$ MeV
can therefore be met -- even if only just -- for the quoted
range in $Q_\nu$ and $K$ assuming the parameters fit superburst
ignition conditions in Fig.~\ref{fig1}. Although such fine tuning in
the binding energy can not be ruled out it is rather conspicuous, but
given the simplicity of our model and that we have not dealt with the
uncertainties or the time dependence in detail such quantitative
conclusions should be treated with caution. We have attempted to test
for consistency, not to determine the quark matter parameters. This
would require more detailed modelling, possibly taking the individual
accretion history of each source into account, which may alleviate the
need to keep $Q_\mathrm{SQM}$ small.

Thus even assuming a high $Q_\mathrm{SQM}$~the model seems consistent
for the systems where it should be most credible and with the
exception of\object{ SAX J1808.4-3658} no clear inconsistency can be
demonstrated for the other sources studied here. However \object{SAX
J1808.4-3658} is difficult to explain with neutron stars as well, and
its precise nature remains to be decided. At this point strange stars
thus seem to do no worse than neutron stars in explaining the
quiescent emission of soft X-ray transients regardless of the assumed
core parameters, and for a low $Q_\mathrm{SQM}$ they explain even the
coldest sources including SAX J1808.4-3658. Because the two cooling
mechanisms are important in different accretion regimes it finally
seems likely that detailed modelling along with more accurate
temperature measurements at low accretion rates would constrain the
core parameters further -- or perhaps demonstrate a clear
inconsistency.

Finally we should not omit a short discussion of the interesting quasi
persistent transients KS 1731-260 and MXB 1659-29 \citep{Wijnands:2001,Rutledge:2002b,Wijnands:2002,Wijnands:2003,Wijnands:2004,Cackett:2006}.
These are systems with prolonged accretion outbursts -- 12.5 and 2.5
years duration respectively -- and a convincing case has been built
for the suggestion that their crusts are significantly heated during
these outbursts and brought out of thermal equilibrium with the
core. The thermal components in their spectrum therefore show the
cooling of the crust to thermal equilibrium and recently
\cite{Cackett:2006} using new {\it Chandra} and {\it XMM-Newton} data
found that both systems are now close to equilibrium allowing the core
temperatures and thermal cooling time scales of the crusts to be
determined. Fitting the cooling curves with an exponential decay
towards a constant temperature presumably reflecting the core
temperature the constant was found to be $71.3\pm 1.6$ eV for KS
1731-260 and $52.1\pm 1.6$ eV for MXB 1659-29 at infinity. Using a
canonical mass-radius relationship this implies surface temperatures
of $93.1\pm 2.1$ eV and $68.0\pm 2.1$ eV respectively, so these
sources are both cold but not unusually so. Only upper limits to
the accretion rates can be established with any certainty however due
to the very limited number of observed outbursts. For KS 1731-260 (one
outburst in 1988-2001) this is $5.1\times 10^{-9} \mbox{
M}_\odot\mbox{ yr}^{-1}$ \citep{Yakovlev:2003} using the time-averaged
flux of \cite{Rutledge:2002b}. For MXB 1659-29 \cite{Wijnands:2003}
find bolometric fluxes of $5-10\times 10^{-10} \mbox{ erg
cm}^{-2}\mbox{ s}^{-1}$ for the 1999-2001 outburst and $2-12\times
10^{-10} \mbox{ erg cm}^{-2}\mbox{ s}^{-1}$ for the 1976-1979 outburst
which using the highest of these values in Eq.~\eqref{mdot} (with
$f=1$) means an upper limit of $1.6\times
10^{-10}\mbox{ M}_\odot\mbox{ yr}^{-1}$ for a distance of 10
kpc. Referring to Fig.~\ref{fig3} both of these limits are very high
given the surface temperatures, and would require enhanced cooling for
both strange stars and neutron stars -- but then the accretion rates
are hardly certain enough to back up such a claim. Perhaps more
interesting the $e$-folding times for the crust temperature relaxing
back to the core temperature were determined both sources. Both of
these were rather short ($246 \pm 62$ days for KS 1731-260 and $501\pm
61$ days for MXB 1656-29) and by comparison with the neutron star cooling
curves computed by \cite{Rutledge:2002b} for the accretion history of
KS 1731-260 may be taken as evidence for high thermal
conductivity in the crust. Strange stars are missing the deep crust
between $10^{11}$ and $10^{15}\mbox{ g cm}^{-3}$ and as we have seen
there is very little tunnelling in quiescence, so although it would
go too far to show this in detail here, it should be safe to assume that
they cool faster than neutron stars relaxing the need for high
thermal conductivity in the crust. This new development therefore
further underlines how rewarding a time dependent model for accreting
strange stars might be.

\acknowledgement{This work was supported by The Danish Natural Science Research Council.}

\end{document}